# The Kepler-SEP Mission: Harvesting the South Ecliptic Pole large-amplitude variables with Kepler

R. Szabó[1], L. Molnár[1], Z. Kołaczkowski[2], P. Moskalik[3], Ž. Ivezić[*,1,4], A. Udalski[5], L. Szabados[1], C. Kuehn[6], R. Smolec[3], A. Pigulski[3], T. Bedding[6], C. C. Ngeow[7], J. A. Guzik[8], J. Ostrowski[2], P. De Cat[9], V. Antoci[10], T. Borkovits[11,12], I. Soszyński[5], R. Poleski[5], Sz. Kozłowski[5], P. Pietrukowicz[5], J. Skowron[5], D. Szczygieł[5], Ł. Wyrzykowski[5], M. Szymański[5], G. Pietrzyński[5], K. Ulaczyk[5], E. Plachy[1], J. Schou[13], N. R. Evans[14], G. Kopaczki[2]

[1]*Konkoly Observatory, Research Center for Astronomy and Earth Sciences of the Hungarian Academy of Sciences, H-1121 Konkoly Thege Miklós út 15-17, Budapest, Hungary*
[2]*Instytut Astronomiczny Uniwersytetu Wrocławskiego, Kopernika 11, 51-622 Wroclaw, Poland*
[3]*Nicolaus Copernicus Astronomical Centre, ul. Bartycka 18, 00-716 Warsawa, Poland*
[4]*University of Washington, Department of Astronomy, P.O. Box 351580, Seattle, WA 98195-1580, USA*
[5]*The OGLE Collaboration: Warsaw University Observatory, Aleje Ujazdowskie 4, 00-478 Warsawa, Poland*
[6]*Sydney Institute for Astronomy (SIfA), University of Sydney, NSW 2006, Australia*
[7]*Graduate Institute of Astronomy, National Central University, Jhongli 32001, Taiwan*
[8]*Los Alamos National Laboratory, Los Alamos, NM 87545 USA*
[9]*Royal observatory of Belgium, Ringlaan 3, B-1180 Brussel, Belgium*
[10]*Stellar Astrophysics Centre, Department of Physics and Astronomy, Aarhus University, Ny Munkegade 120, DK-8000 Aarhus C, Denmark*
[11]*Baja Astronomical Observatory, H-6500 Baja, Szegedi út, Kt. 766, Hungary*
[12]*ELTE Gothard-Lendület Research Group, H-9700 Szombathely, Szent Imre herceg út 112, Hungary*
[13]*Max Planck Institute for Solar System Research, Max-Planck-Str. 2, 37191 Katlenburg-Lindau, Germany*
[14]*Harvard-Smithsonian Astrophysical Observatory, MS 4, 60 Garden St., Cambridge, MA 02138, USA*
[*]*Distinguished guest professor invited by the Hungarian Academy of Sciences*

**Abstract**

As a response to the Kepler white paper call, we propose to **turn Kepler to the South Ecliptic Pole and observe thousands of large amplitude pulsating and eclipsing variables** for years with high cadence in the frame of the Kepler-SEP (Kepler - South Ecliptic Pole) Mission. The degraded pointing stability will still allow observing these stars with reasonable (probably better than millimag) accuracy. Long-term continuous monitoring already proved to be extremely helpful to investigate several areas of stellar astrophysics, like the century-old Blazhko-enigma. Space-based photometric missions opened a new window to the intricate dynamics of pulsation in several class of pulsating variable stars and facilitated very detailed studies of eclipsing binaries. **The main aim of this mission is to better understand the fascinating dynamics behind various stellar pulsational phenomena** (resonances, mode coupling, period-doubling, chaos, mode selection) and interior physics (turbulent convection, opacities). This will also **improve the applicability of these astrophysical tools for distance measurements, population and stellar evolution studies.** We investigated the pragmatic details of such a mission and found a number of advantages: minimal reprogramming of the flight software, a favorable field of view, access to both galactic and LMC objects. However, **the main advantage of the SEP field comes from the largest possible sample of well classified targets, mainly through OGLE**. Synergies and significant overlap (spatial, temporal and in brightness) with both ground-based (OGLE, LSST) and space-based missions like GAIA and TESS will greatly enhance the scientific value of the Kepler-SEP mission. GAIA will allow full characterization of the distance indicators (calibration of the zero point of the period-luminosity diagram), by providing distances. TESS will continuously monitor this field for at least one year, and together with the re-purposed Kepler mission provide long time series data that cannot be obtained by other means. If Kepler-SEP program is successful, there is a possibility to place one of the so-called LSST "deep-drilling" fields in this region.

## 1. Introduction

In July 2012 one of the four Kepler reaction wheels failed, while in May 2013 a second one showed elevated friction levels prompting NASA engineers to initiate immediate actions to preserve the functionality of this unique spacecraft. Indeed, the Kepler space telescope (Borucki et al. 2010) opened a new window to exoplanetary systems (Batalha et al. 2013), as well as allowed an unprecedented view on stars (see e.g. Gilliland et al. 2010 and Bedding et al. 2011). After unsuccessful recovery attempts a white paper call[1] was initiated to allow experts from any fields of astrophysics to propose new missions exploiting the capabilities of Kepler handicapped by reduced pointing accuracy. We propose to **turn Kepler to the South Ecliptic Pole and observe many thousands of large amplitude pulsating and eclipsing variables** in this single field with high cadence for years in the frame of the Kepler-SEP (Kepler - South Ecliptic Pole) Mission. Kepler's field-of-view can cover large part of the LMC halo and part of the crowded LMC bar (Fig. 1.).

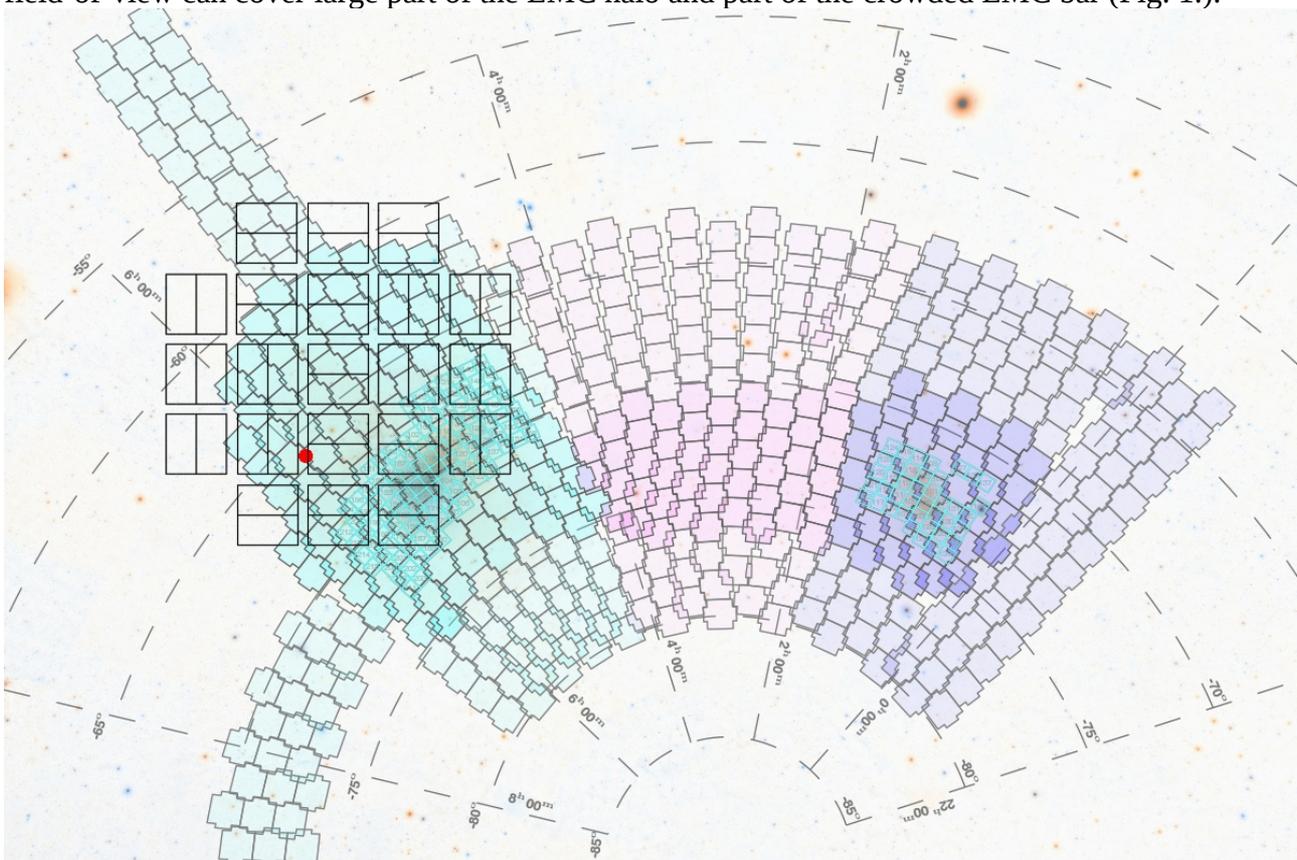

*Figure 1. The relation between the OGLE-IV coverage and the Kepler field of view. The space telescope can observe the region around the Southern Ecliptic Pole (red dot) and a large part of the Large Magellanic Cloud. The OGLE-IV survey can be easily extended to cover the missing part from the Kepler-SEP field.*

## 2. Science goals

In this section we summarize the scientific motivations to observe high-amplitude variable stars in the Southern Ecliptic Pole field by variable star classes.

### 2.1. *RR Lyrae stars*

Relatively small space telescopes delivered new results on these high-amplitude pulsating stars. The Canadian MOST (Microvariability and Oscillations of STars) telescope for example monitored a double-mode RR Lyrae star, AQ Leo (Gruberbauer et al. 2007) establishing the existence of a new periodicity most probably originating from the presence of nonradial modes. CoRoT (Convection, ROtation and planetary Transit) continued the exploration and established new standards in the observations of Blazhko-modulated RR Lyrae stars (Chadid et al. 2010, Guggenberger et al. 2011). The Blazhko-effect is the mysterious amplitude and phase modulation of RR Lyrae stars. The unprecedented Kepler Mission crowned the monitoring of the Blazhko stars (Benkő et al. 2010), unveiling new and unexpected dynamical phenomena, like the period doubling (Szabó et al. 2010), unexpected triple-mode state (Molnár et al. 2012) and even low-dimensional chaos (Plachy et al. 2013). These new discoveries are tied strongly to the Blazhko-enigma and induced new types of theoretical studies (Kolláth et al. 2011, Smolec & Moskalik 2012) that in turn led to a new explanation of the Blazhko effect by a resonance between the radial fundamental mode and a high overtone pulsation mode (Buchler & Kolláth 2011).

---

[1] http://keplerscience.arc.nasa.gov/docs/Kepler-2wheels-call-1.pdf

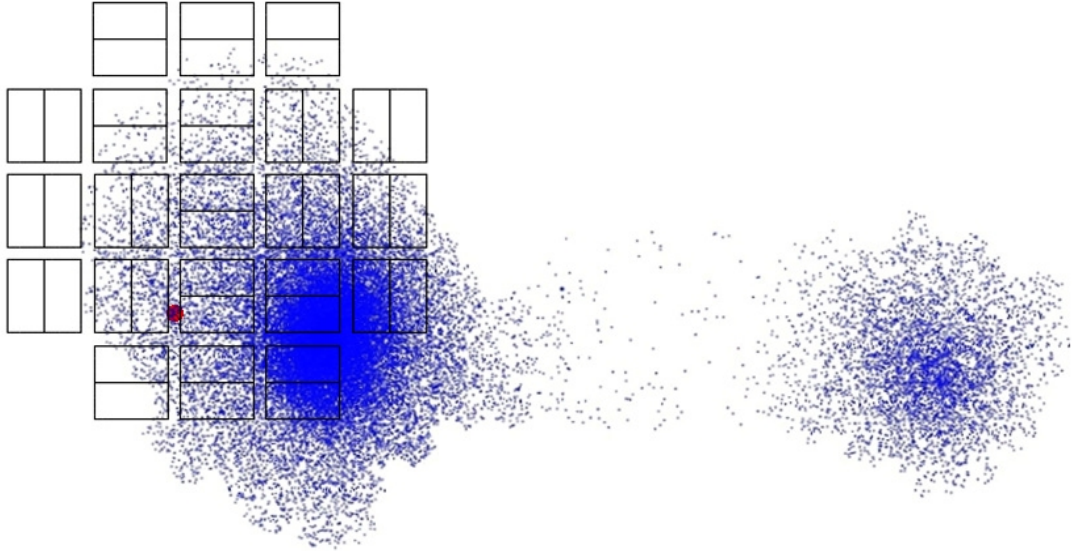

*Figure 2. The distribution of RR Lyrae stars in and around the LMC (left) and SMC (right) based on the OGLE observations. Note the large number of RR Lyrae variables in the vicinity of the South Ecliptic Pole (denoted by a red circle).*

Kepler observed about 40 RR Lyrae stars during its primary mission. In contrast, the SEP fields contain thousands of RR Lyrae located in the Milky Way halo and the LMC (Fig. 2.). Although the LMC RR Lyrae are quite faint, between magnitudes 18-19 in Johnson V color, Kepler has already observed stars this faint in the primary field with satisfactory results. Kepler will easily provide better estimates on the frequency and variety of the Blazhko-effect in different galactic environments than any previous study. Continuous high-cadence observations will also reveal if the dynamical phenomena Kepler has discovered in field stars are present in those populations as well (Benkő et al. 2010, Szabó et al. 2010, Kolláth et al. 2011, Molnár et al, 2012) and may lead to the explanation of the century-old mystery (Blazhko 1907).

Several hundreds of the RR Lyrae variables are first overtone (RRc) or double-mode (RRd) stars. Space-based photometry exists only for a handful of such stars so far, hence the Kepler-SEP mission will have a huge potential for the understanding of these stars as well. In particular, the frequency of modulation among RRc stars (first overtone pulsators) and the presence of additional, low-amplitude modes in all RR Lyrae classes are of great interest (Moskalik et al. 2013).

*2.2. Cepheids*

In the original Kepler field only one classical Cepheid has been found (Szabó et al. 2011). We will be able to follow at least two orders of magnitude more targets of this class (both fundamental and first overtone Cepheids, as well as different double-(or even) triple mode pulsators) following the present proposal (Fig. 3.). Several Cepheids display pulsation properties that are poorly understood. Some stars show periodic amplitude and phase variations in two modes simultaneously, similar to the Blazhko-effect in RR Lyrae stars.

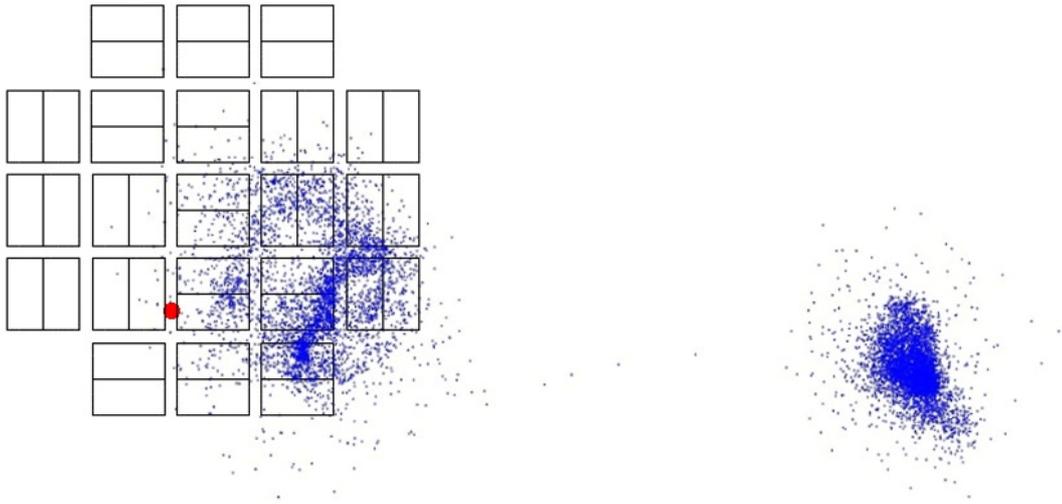

*Figure 3. Distribution of Cepheids in the LMC (left) and SMC (right) regions based on the OGLE observations. The bar of LMC is clearly seen. Hundreds of Cepheids are located in the SEP region. The proposed Kepler field-of-view and the South Ecliptic Pole (red dot) are overplotted.*

Additional, low-amplitude modes are observed in several stars, some of which can be explained by radial modes (e.g. Soszynski et al. 2008), and others that appear to be non-radial modes but have similar characteristics in many stars (e.g. Moskalik & Kolaczkowski 2009, Dziembowski 2012). Strange modes, high-order, low-amplitude radial modes were also predicted to exist in Cepheids, yet there are no confirmed observations so far (Buchler & Kolláth 2001). Kepler is perfectly suited to detect and follow these variations and to advance their understanding.

Cepheids in eclipsing binaries are rare but extremely valuable stars as pulsations and orbital dynamics provide simultaneous yet independent estimates of fundamental stellar parameters (Pietrzynski et al. 2010). One such system was already detected in the GAIA SEP field, and more could be hiding among the hundreds of Cepheids Kepler will observe. Due to its better sampling and smaller photometric errors than ground-based observations Kepler will be capable of finding many more such objects with much shallower eclipses.

The detailed study of the Kepler data of the only bona fide Cepheid in the Kepler field clearly indicated that the pulsation of V1154 Cygni is not strictly periodic (Derekas et al., 2012). Such behavior was expected judging from the excess scatter of the ground-based light curves of large number of Cepheid pulsating in the first overtone (Klagyivik & Szabados, 2009). The SEP field contains more than 40 Cepheids pulsating in the first overtone, and a larger number of fundamental-mode Cepheids. The continuous photometric coverage of the SEP region helps to reveal whether the period jitter is a common property of overtone Cepheids, and would allow us a comparison of the light curve stability of fundamental- and first-overtone Cepheids. Is such fluctuation affected by metallicity? LMC Cepheids observed by Kepler-SEP will help to answer. Any significant result would have an implication on the pulsation theory. Cepheids in the LMC are important steps in the extragalactic distance scale ladder, therefore a better understanding of their pulsational properties is also desirable. A number of LMC Cepheids covered in Kepler-SEP fields have, or will have, multiband data available, such as in JHK from (Persson et al. 2004) or VMC, and MIR photometry from Spitzer SAGE (Scrowcroft et al 2011). Of course, extensive V and I band light curves will also be available from OGLE-III and OGLE-IV.

### 2.3. Type II Cepheids

There are five Type II Cepheids and three anomalous Cepheids in the SEP field. These types of variables have never been observed from space, therefore lack high-precision light curves. From the available ground-based photometric data it is known that (i) the pulsation period of Type II Cepheids is much less stable than that of classical Cepheids (ii) the anomalous Cepheid XZ Ceti has a slightly unstable photometric phase curve (Szabados et al., 2007). The continuous coverage of the brightness variability of the SEP field would allow us to study the cycle-to-cycle variations of these neglected types of Cepheids and unveil the nature of the dynamics of their pulsation.

### 2.4. A-F pulsators on the main sequence

Delta Scuti stars in the Magellanic Clouds are faint: Kepler may provide useful photometry of only of the brightest handful of them. The catalog by (Poleski et al. 2010) contains 19 delta Sct stars in the foreground of the OGLE-III LMC fields. They are bright enough for Kepler to study. We may expect ~10 to be present in Kepler SEP field. Because of their fast variability, with periods of a few hours, however, Kepler short cadence observations will still be superior to the sparse OGLE data. On the other hand, the drift noise will affect the low-frequency signals more significantly, making the detection of gamma Doradus and hybrid stars difficult, but not impossible.

The delta Scuti as well as the gamma Dor stars are important astrophysical objects for stellar structure studies because they occupy the region in the HRD where the transition between deep and efficient to shallow convective envelopes takes place. The so-called granulation boundary cuts right through the δ Sct and γ Dor instability strips (Gray & Nagel 1989). Recent studies suggest that convective motions are still present in mid to late A type stars (Landstreet et al. 2009, Kallinger & Matthews 2010). Because convection has an impact not only on pulsational stability but also on stellar evolution, activity, modelling stellar atmospheres, transport of angular momentum, etc., it is of great interest to understand how this transition takes place.

Stars later than F5 have very slow rotation rates, with a sharp increase between F0 and F5 (Royer 2009). The measured rotational velocities for the cool stars inside the instability strip are on average higher than 100 kms$^{-1}$. Royer (2009) also reports observational evidence for differential rotation in A type stars, which is related to magnetic dynamos like in the Sun. Chromospheric activity disappears only for stars hotter than 8300 K (Simon et al. 2002), demonstrating that the instability strip of δ Scuti and γ Doradus pulsators are affected by this transition.

While straight-forward modelling of individual frequencies is complicated, because we have little information on the geometry of pulsation modes, studying the stability of these oscillations can still help to better characterize the pulsation mechanisms. From recent observations from Kepler we now know that the current theoretical understanding is either missing basic physics or that an additional excitation mechanism must be taken into account (Antoci, private communication). Long uninterrupted observations, even of degraded quality, are crucial for the aforementioned studies.

### 2.5. B-type pulsators

Pulsating B stars were essentially missing form the original Kepler field. LMC has large populations of early B-type stars that fall well within the appropriate magnitude range. If magnitude 18.5 is selected as a limit, this corresponds to B8 on the main sequence (Pigulski & Kołaczkowski 2002; Kołaczkowski et al., 2006), allowing monitoring of thousands of B stars. Pulsating B stars (beta Cep and SPBs) are excellent test objects for internal physics like convective overshooting (Degroote et al. 2010). Moreover, a detailed study of the B-type star pulsators may also lead to clues to solve the metallicity problem. Theoretical models predict that the beta Cephei-type and SPB-type pulsations should vanish for metallicities Z<0.01 and Z<0.006, respectively (Pamyatnykh, 1999) while the LMC ($Z_{LMC}$=0.008) has several known pulsators from both types (Kolaczkowski et al., 2006). Kepler-SEP might be able to find members of a new and poorly known pulsating variable star class, the Slowly Pulsating B-type supergiants (Saio, H., et al. 2006, Daszynska-Daszkiewicz et al. 2013). Massive binary stars also present a great opportunity for the Kepler-SEP mission.

### 2.6. *Eclipsing binaries, exoplanets*

Eclipsing binaries are the Swiss knifes of stellar astrophysics that may provide fundamental stellar parameters and distances with very little need of assumptions and calibrations. Long-term, high-cadence, precise monitoring may also reveal dynamical effects in multiple systems, as it did already in the primary mission (Slawson et al. 2011). Eclipse-timing variations (ETVs), tertiary eclipses and changing eclipse depths will still be detectable in many binary systems, despite the degraded photometric accuracy. In the majority of the known ETV binaries the eclipse depth exceeds one percent (feasible with Kepler-SEP), while one third of these systems show long-term ETVs, which necessitates years-long observations. The Kepler-SEP mission may also be able to discover new exoplanets, even circumbinary ones through eclipse-timing variations in the same way as tertiary stars in these systems (Sybilski et al. 2010, Konacki 2012).

### 2.7. *Other variable stars*

Beyond the classes above, Kepler will be able to study a very diverse collection of objects. The OGLE-IV classification scheme (Soszynski et al. 2012) for the GAIA SEP field includes two large groups (Long-Period Variables and Other) that contains stars of very different classes: non-periodic variables, Be and spotted stars and cataclysmic variables. These groups may represent a treasure trove of stellar astrophysics. In addition, there might be a few intrinsically faint variables among the foreground stars, like pulsating white dwarfs and subdwarfs.

A few supernovae that are bright enough may be available for Kepler in the SEP field each year. If the telescope can be alerted fast enough, it may be able to follow the bright phase of the eruptions. It would be a unique opportunity since to our knowledge no supernova has been observed quasi-continuously with the available Kepler precision.

### 2.8. *Galaxy formation and population studies*

We stress that an additional bonus of this proposal is the access to different galactic and LMC populations in the SEP field. This ensures that the full pulsational characterization of the variable star samples connected to population and kinematic studies will provide a unique laboratory to test galactic formation theories and to study the history of our own Galaxy (e.g. Miglio et al., 2009, Sesar et al. 2013). Different classes of variable stars correspond to stellar populations of different ages. Classical Cepheids trace recent star formation while RR Lyrae stars trace oldest populations in halo streams, remnants of merging dwarf spheroidal galaxies, globular cluster tidal tails, and other early structures of galaxy formation.

3. **Synergies with OGLE, GAIA, TESS and LSST**

Thanks to the OGLE survey we have an exhaustive knowledge on the variable star populations in the Kepler-SEP field. The OGLE-IV survey can be easily extended to cover the missing part from our targeted field (Fig 1). **The SEP field will be extensively monitored by other missions (GAIA, TESS, LSST).** Synergies with (OGLE) and these space-based missions will enhance the scientific value of such a mission.

GAIA will allow a full characterization of the distance indicators (calibrating the zero point of the period-luminosity relationship), by providing distances. During the commissioning phase of GAIA, a special scanning mode repeatedly covering the ecliptic poles on every spin is planned for calibration purposes. These data cover small, 1-degree fields around the ecliptic poles with high redundancy. The GAIA ecliptic-pole data will be released two years after launch[2].

The Kepler-SEP mission can be viewed as a precursor mission to TESS, since TESS will be able to continuously monitor this field for at least one year, and together with the re-purposed Kepler mission provide long time series data that cannot be obtained by other means.

---

[2]http://www.rssd.esa.int/index.php?project=GAIA&page=Data_Releases

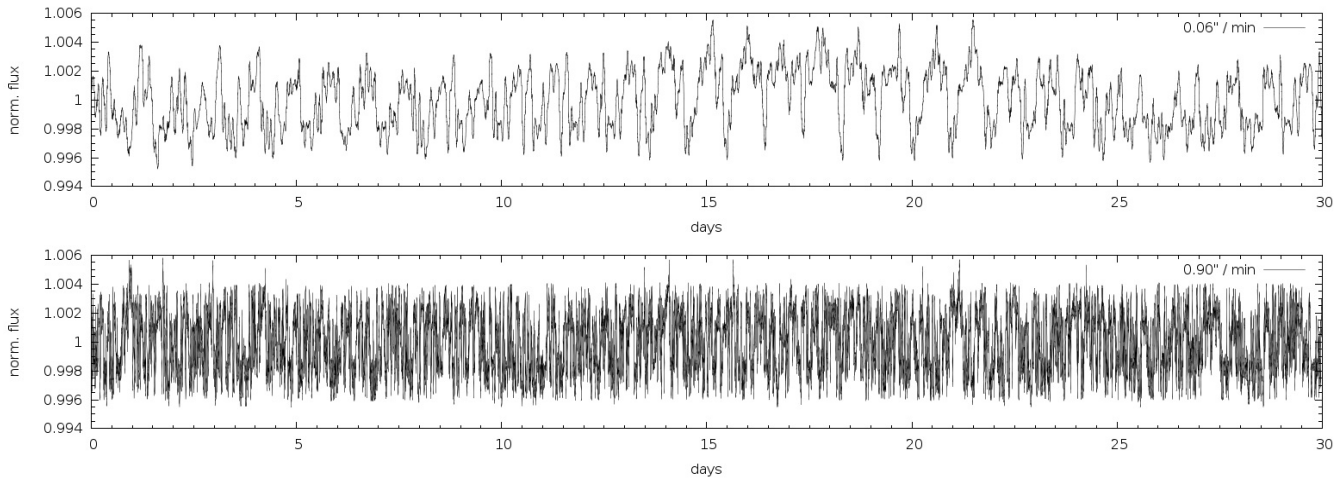

*Figure 4. The estimated level and shape of the drift noise, based on the Kepler photometry simulator, with two different drift rates. Calculated by Kjeldsen, Arentoft & Christensen-Dalsgaard (2013).*

This region will also be observed by LSST – one of the key projects identified in the 2010 Astronomy and Astrophysics Decadal Survey – about 1000 times (summed over *ugrizy* passbands). The LSST saturation limit (r ~16 and brighter in other passbands) is bright enough to provide substantial overlap with the program proposed here. In addition, if this program is successful, there is a possibility to place one of the so-called LSST "deep-drilling" fields (with field-of-view area of 9.6 sq. deg.) in this region, which could provide substantially denser temporal coverage. Since LSST will operate until 2032, together with all other data, this region would have by far the best temporal coverage for studying long time-scale phenomena.
**Kepler has a unique aspect to add to these synergies – the short cadence and continuous monitoring during 2-4 years of the Kepler-SEP mission lifetime.**

4. **Observing strategy, field and mission length**

We propose the region around the Southern Ecliptic Pole that covers the continuous viewing zones of future space-based missions as a hunting ground for the resurrected Kepler. The exact position will depend on several factors: optimal spacecraft attitude, target selection, crowding towards the LMC. Based on the available information of the telescope's current status, we designed the Kepler-SEP mission with a lifetime of four years. To achieve the science goals listed in Sec. 2. a minimum operation of 1 year is required.

4.1. **Pointing**

According to the description in the Call for White Papers, the South Ecliptic Pole is a highly favorable direction and the telescope may observe it continuously throughout the year. In this case the pointing is oriented perpendicular to the orbital plane (case B in the Explanatory Appendix for the Call) and the boresight error can be limited to a few pixels with 1-day pointing attitudes and daily corrective thruster firings. A strong limitation of the perpendicular attitude is the rotating field-of-view: in order to keep the spacecraft orientation close to the balance point, the boresight will have to be rotated by 1 degree at each daily pointing tweak.

4.2. **Expected photometric accuracy**

A detailed simulation of the expected photometric accuracy was carried out within the KASC[3]. Global pixel-to-pixel variations, intrapixel variations and sensitivity drops due to pixel-channels based on the Kepler Instrument Handbook were incorporated into the Kepler photometry simulator (De Ridder, Arentoft & Kjeldsen, 2006). According to the results, the drift introduces variations on the levels of ± 4000 ppm with characteristics that highly depend on the actual rate and amount of the drift (Figure 4). This drift noise is too high to detect the intricate solar-like oscillation signals but other stars with higher amplitudes and/or time scales that differ sufficiently from the characteristics of the noise should be accessible. Large-amplitude variability, either slow or fast is mostly unaffected by the drift except for the finest details in the light curves (Fig. 5). Depending on the actual drift rate, low-amplitude, high-frequency variations (e.g. pulsations in white dwarfs and delta Scuti variables) may be also preserved.

---
[3] Kepler Asteroseismic Science Consortium, http://astro.phys.au.dk/KASC/

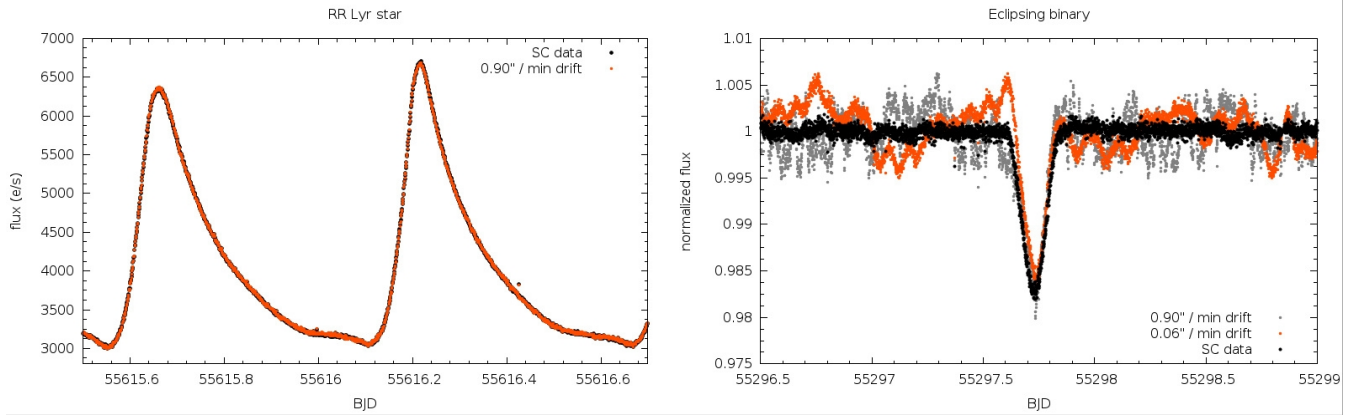

*Figure 5. Effects of the drift on different types of variables. RR Lyrae stars (left), thanks to their large amplitudes, will be relatively insensitive to the drift noise. Eclipsing binaries will be more vulnerable: the right panel shows a star where the length (~0.2 d) and depth (1.8%) of the eclipse is affected by the slower simulated drift rate (orange points).*

Photometric accuracy will be further affected by the roll of the field-of-view as stars may move onto different modules over time. However, these estimates are still very preliminary and the actual noise levels will strongly depend on the telescope performance and post-processing capabilities. Simultaneous observations by OGLE may also aid the removal of systematics from the data. **We expect that the noise can be limited to ~1000 ppm or lower, especially close to the center of the field-of-view.**

### 4.3.  **Targets**

One of the advantages of the proposed field is that it has been and are being observed by various programs. The OGLE-IV project has surveyed the SEP since 2010, providing a very detailed and homogeneous data set. **The OGLE and MACHO databases ensure that the targets may be selected with reasonable effort, without the need of extensive additional observations.**

Although the entire data is not available yet, results for the four OGLE fields covering 5.3 square degrees around the SEP have been published (Soszynski et al. 2012). Cepheids are intrinsically bright (13-16 apparent magnitude at the distance of LMC) and excellently suited for photometric follow-up with the limited-precision Kepler photometry. Some stars, like RR Lyrae stars and delta Scuti stars are quite faint. If we limit the targets to 18.5 magnitudes, the following statistics arise (selected versus all) for the OGLE-IV SEP field alone:

- Classical Cepheids: 132 / 132
- Anomalous and type II Cepheids: 5 / 5
- RR Lyrae stars: 96 / 686   (674 / 686 if the limit is 19.0)
- delta Scuti stars: 4 / 159   (  9 / 159 if the limit is 19.0)
- eclipsing binaries and ellipsoidal variables: 560 / 1533
- long-period variables: 2791 / 2819
- others (including spotted and Be stars): 1055 / 1473

Based on these statistics, the entire Kepler-SEP field of view is expected to contain 20 times more pulsating variable stars than the above numbers, though the modules close to or falling onto the LMC would suffer from crowding.

### 4.4. Integration times, masks, and the number of targets

The drift and roll of the telescope requires modifications in the procedure of allocating the masks. The two-arcminute boresight error in attitude translates to only a few pixels at the edges of the field-of-view. The 1-degree rotation on the other hand will require either a few long arcs along the modules or the ability to automatically shift/reposition the masks during the pointing tweaks. Given the large number of possible targets in the field, we favor the latter scenario, understanding that it will require modifications in the flight software.

The typical cadence of OGLE-IV in Magellanic Clouds is once or twice per night with seasonal gaps. Only a few fields located in the most dense part of the Galactic Bulge are observed 10-30 times per night. Faster sampling is one of the main advantages of Kepler that we wish to exploit. Hence we intend to seriously limit the use of long cadence mode. We instead propose to limit the number of stars and observe a few thousand stars at least in short or intermediate cadence mode. If all targets are observed with short cadence (~1-minute), the data transfer rate to the recorder limits the number of saved pixels to 300 000 - 400 000. With the possible requirement of larger masks, that results in a few thousand individual targets. Targets can also be changed as the field-of-view turns.

Alternatively, choosing an intermediate cadence could increase the number of observed targets. Specifically, co-addition for five minutes would allow for about 10 000 - 40 000 targets. A higher number of targets allows to keep a limited number of masks 'vacant' for an alert mode, for example to observe bright supernovae or transits of large (Jupiter- and Neptune-size) planets and variability surveys. To optimize the scientific output the best solution would be to allocate at least half of the photometric masks to high (1-minute) cadence observations and to assign intermediate cadence for the remaining resources.

### 5. Conclusions

Despite the reaction wheel failures, Kepler remains a space telescope of great value. The Kepler-South Ecliptic Pole mission is a great possibility to exploit its potential. The area around the Southern Ecliptic Pole is unparalleled in the whole sky, because it is filled with the largest number of pulsating and eclipsing objects both from our Galaxy and the Large Magellanic Cloud accessible to Kepler. High stellar density due to the proximity of LMC will affect only a limited part of the field-of-view. The mission will answer fundamental questions of stellar astrophysics through high-amplitude pulsating variable stars, binary and multiple star systems and can unveil differences between stellar populations in our Galaxy and the Large Magellanic Cloud.

Synergies with current and future observing programs represent offer great advantages. Kepler will observe in parallel with OGLE-IV, providing high-cadence, uninterrupted observations for the most interesting targets the ground-based survey uncovers. The area covered by OGLE-IV observations is easily extended to encompass the entire Kepler-SEP field. Parallel observations will be useful in post-processing to lower the noise and to remove the systematics from the data.

The South Ecliptic Pole will be monitored by several future projects (GAIA, TESS, LSST) providing the best temporal and spatial coverage on the sky and the contribution of Kepler will be important in this unique cooperation. The Kepler-SEP mission will also serve as a precursor for TESS, extending the observations of bright targets in the southern continuous viewing zone of TESS. But faint targets will also be of great importance to extend the observations of LSST: it is possible to place one of the planned "deep-drilling" fields onto the Kepler-SEP stars.

We believe that the need of only a minimal or modest reprogramming of the flight software, the favorable field of view and the synergies with so many astronomical surveys make the Kepler-SEP mission ready to fill new pages in the virtual book of Kepler's impressive scientific achievements.

**Acknowledgements:**

We gratefully acknowledge the following grants: the Lendület-2009 Young Researchers' Program of the Hungarian Academy of Sciences, the HUMAN MB08C 81013 grant of the MAG Zrt., the Hungarian OTKA grant K83790 and the KTIA URKUT_10-1-2011-0019 grant. We wish to thank organizers of the IAU Symposium 301 (Precision Asteroseismology: Celebrating the scientific opus of Wojtek Dziembowski) for allowing us to organize a splinter session on the future of Kepler. R. Sz., L. M., P. M. and R. S. acknowledge the IAU grants for the conference. R. Sz. acknowledges the János Bolyai Research Scholarship of the Hungarian Academy of Sciences. Ž.I. thanks the Hungarian Academy of Sciences for its support through the Distinguished Guest Professor grant No. E-1109/6/2012. The OGLE project has received funding from the European Research Council under the European Community's Seventh Framework Programme (FP7/2007-2013) / ERC grant agreement no. 246678 to AU.